\documentclass[10pt]{iopart} 
\usepackage{iopams}

\usepackage{graphicx}
\usepackage{dcolumn}
\usepackage{bm}

\usepackage{color} 
\usepackage{ulem} 
\usepackage{siunitx} 

\usepackage{textcomp} 

\begin{document}

\title{Superconductivity in Ternary Germanite TaAl$_{x}$Ge$_{2-x}$ with a C40 chiral structure}

\author{
Daigo Koizumi$^{1}$,
Shingo Kisanuki$^{1}$,
Kenta Monden$^{1}$,
Yusuke Kousaka$^{1*}$,
Hiroaki Shishido$^{1}$,
Yoshihiko Togawa$^{1}$
}
\address{$^{1}$ Department of Physics and Electronics, Osaka Metropolitan University, Sakai, Osaka 599-8531, Japan}

\ead{koyu@omu.ac.jp}

\vspace{10pt}
\begin{indented}
\item[]October 2024
\end{indented}

\begin{abstract}
We report a new family of chiral intermetallic superconductors TaAl$_{x}$Ge$_{2-x}$. The mother compound TaGe$_2$ has a C40-type chiral hexagonal crystal structure with a pair of enantiomorphic space groups of $P6{_2}22$ and $P6{_4}22$.
By substituting Ge with Al, TaAl$_{x}$Ge$_{2-x}$ polycrystals with the C40 structure were synthesized with Al substitution $x$ from 0 to 0.8. Magnetic susceptibility, magnetization curves and electrical resistivity revealed that TaAl$_{x}$Ge$_{2-x}$ with $x$ of 0.2 to 0.4 was a type-II superconductor with a superconducting transition temperature $T_{\rm c}$ of 2.0 to \SI{2.2}{\kelvin}.
The superconductivity disappeared or was largely suppressed at $x$ less than 0.2  and more than 0.4,
although all the measurements were performed at temperatures above \SI{1.8}{\kelvin}. An emergence of superconductivity is discussed in terms of the lattice constants changes with the Al substitution.
\end{abstract}

%
%
%
%
\ioptwocol

\section{Introduction}

Superconductors with a non-centrosymmetric crystal structure have stimulated the search for materials with unconventional Cooper pair states and enhancing properties of superconductivity. 
Indeed, an antisymmetric spin-orbit coupling (ASOC) due to a lack of inversion symmetry leads to a mixture of spin-singlet and spin-triplet states. 
Such a mixed pairing state allows an extremely high upper critical field that exceeds the Pauli paramagnetic limit, as observed in CePt$_{3}$Si \cite{Bauer2004,Yasuda2004}.

In this connection, superconductors with a chiral crystal structure (chiral superconductors), 
which lacks horizontal and vertical mirrors as well as an inversion center, can be classified as a part of non-centrosymmetric superconductors. 
For instance, Li$_{2}$Pd$_{3}$B and Li$_{2}$Pt$_{3}$B, which are one of the representative 
chiral superconductors with a cubic crystal structure,
have been investigated in terms of  a weight shift of the spin-triplet component in the mixed pairing state \cite{Togano2004,Badica2005,Yuan2006}.

In addition, chiral crystals are found to exhibit a variety of nontrivial physical phenomena,
as exemplified by nonreciprocal magneto-chiral dichroism (MChD) \cite{Rikken1997,Kibayashi2014,Okumura2015,Nakagawa2017},
electrical magneto-chiral anisotropy (eMChA) \cite{Rikken2001,Yokouchi2017,Aoki2019}
and chirality-induced spin selectivity (CISS) \cite{Gohler2011,Xie2011,Inui2020,Kousaka2023,Shiota2021,Shishido2021,Shishido2023,Ohe2024}.
In these phenomena, the signal sign depends on the handedness of the chiral crystal structure.
Namely, the dynamical physical response has a strong correlation with structural chirality \cite{Togawa2023}.
For example, chiral organic superconductors show a chirality-dependent spin accumulation via the CISS in the superconducting transition regime \cite{Nakajima2023}. 
In this respect, chiral superconductors could potentially have unique attractive properties, which have not been studied well in non-centrosymmetric superconductors.

Very few number of chiral superconductors have been reported so far. 
Most of them form a cubic structure, although monoaxial chiral crystals are in favor of having simplified forms of the ASOC and intensified signals rather than cubic crystals. 
Non-cubic chiral superconductors are limited to the following compounds:
UIr \cite{Akazawa2004}, BiPd \cite{Joshi2011}, 
$T$Rh$_{2}$B$_{2}$ ($T$ = Nb, Ta) \cite{Carnicom2018},
NbGe$_{2}$ \cite{Remeika1978}, TaSi$_{2}$ \cite{Gottlieb1992} and TaGe$_{2}$ \cite{Knoedler1979}.

\begin{figure}[tb]
\begin{center}
\includegraphics[width=8cm]{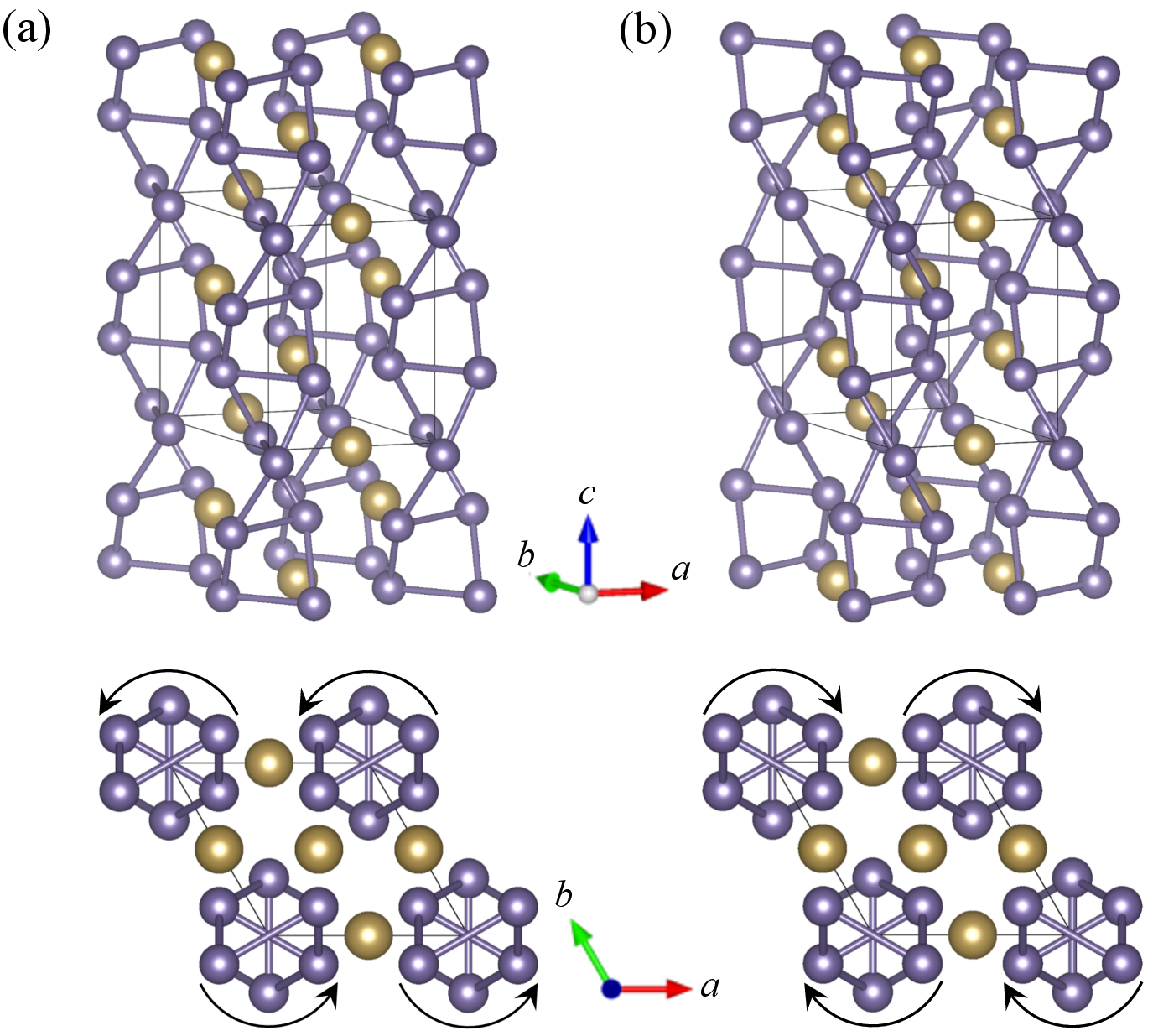}
\end{center}
\caption{Crystal structures of (a) right-handed (space group: $P6_{2}22$) and (b) left-handed (space group: $P6_{4}22$) TaGe$_{2}$
with a C40 chiral structure.
The lower schematics show the (001) projection of each structure.
Large brown and small purple balls represent Ta and Ge atoms, respectively.
The black arrows are given as an eye guide to recognize the sense of screw alignments of the Ge atoms.}
\label{f-TSi-str}
\end{figure}

C40-type disilicides and digermanides $TX_{2}$ ($T$ = Nb, Ta and $X$ = Si, Ge) show hexagonal chiral crystal structures with the enantiomorphic space group of $P6_{2}22$ or $P6_{4}22$. 
Single helix of $T$ and double helices of $X$ form along the hexagonal $c$-axis, as shown in Fig.~\ref{f-TSi-str}. 
The handedness of the helices is determined by the enantiomorphic space groups \cite{Sakamoto2005,Onuki2014}.

TaSi$_{2}$ and NbGe$_{2}$ show superconductivity with a critical temperature $T_{\rm c}$ of 0.35 and \SI{2.0}{\kelvin}, respectively \cite{Remeika1978, Gottlieb1992, Lv2020}.
It was reported that, while TaSi$_{2}$ was a type-I superconductor with the critical magnetic field $H_{\rm c}$ of \SI{29.8} Oe \cite{Gottlieb1992},
NbGe$_{2}$ presented a crossover from type-I to type-II superconductivity below \SI{0.9}{\kelvin} with the lower and upper critical magnetic fields ($H_{\rm c1}$ and $H_{\rm c2}$) of 223 and 360 Oe, respectively \cite{Lv2020}.
TaGe$_{2}$ exhibited zero electrical resistance as an indication of superconductivity at 1.9 and \SI{2.7}{\kelvin} in bulk crystals and thin films, respectively \cite{Knoedler1979}.
Note that the superconductivity was not reproducible in the bulked TaGe$_{2}$ samples with no information available on the superconducting volume fraction (VF) because of a lack of susceptibility data.

In this study, we reexamined superconductivity in C40-type TaGe$_{2}$ and its family compounds. 
New chiral superconductors
were found while substituting Ge with Al for increasing the density of state (DOS) on the Fermi surface.
In magnetization and electrical resistivity measurements,
TaAl$_{x}$Ge$_{2-x}$ compounds showed superconductivity at 2.0 to \SI{2.2}{\kelvin} for $x$ of 0.2 to 0.4, respectively. 
Here, $x$ is a nominal substitution amount of Al. 
A type-II nature of superconductivity was found in TaAl$_{x}$Ge$_{2-x}$ in the magnetization curves. 
On the other hand, TaGe$_2$ did not indicate any sign of superconductivity down to \SI{1.8}{\kelvin}.

\section{Experimental Methods}

Polycrystalline powder specimens of TaAl$_{x}$Ge$_{2-x}$ were synthesized by a solid-state reaction method.
Powders of Ta, Al, and Ge, homogeneously mixed in a molar ratio of 1 : $x$ : $2-x$, were pressed into a pellet.
The pellet was sealed in an evacuated silica tube and prebaked in an electric tube furnace at \SI{1100}{\degreeCelsius} for two days.
Then, it was pulverized and pressed again into a pellet, followed by the same heat treatment to obtain polycrystals of TaAl$_{x}$Ge$_{2-x}$.
In this study, several specimens with $x$ from 0 to 0.8 were prepared to examine an emergence of superconductivity in a series of substituted compounds.

Powder X-ray diffraction experiments were performed with the synthesized powders for phase identification by using a Rigaku Smart-Lab equipment with Cu-K$\alpha$ radiation.
Powder X-ray diffractograms were analyzed by Match! software (Crystal Impact) for the phase identification.
FullProf Suite implemented in Match! was used for refining lattice parameters of TaAl$_{x}$Ge$_{2-x}$ and evaluating an amount of impurity phases.

Magnetization measurements were performed using a superconducting quantum interference device (SQUID) magnetometer (Quantum Design MPMS3).
A magnetic susceptibility was collected as a function of temperature down to \SI{1.8}{\kelvin} at a constant magnetic field of 10 Oe in zero-field cooling (ZFC) and field-cooling (FC) processes.
Magnetization curves were obtained in cycling magnetic fields at different temperatures.
Electrical resistivity was measured down to \SI{1.9}{\kelvin} at a fixed current of \SI{3}{\milli \ampere} with a standard DC four-probe method
using physical property measurement system 
(Quantum Design PPMS).
For each $x$ value, the sample with a size of $8.0 \times 5.0 \times 1.5$ \si{\milli \metre}$^{3}$ was prepared from the synthesized pellets for the resistivity measurements,
while the pulverized powder specimen with a weight of \SI{50}{\milli \gram}, wrapped with a paraffin paper, was fixed in a plastic straw for the magnetization measurements.

\section{Experimental Results}

\begin{figure}[tb]
\begin{center}
\includegraphics[width=8cm]{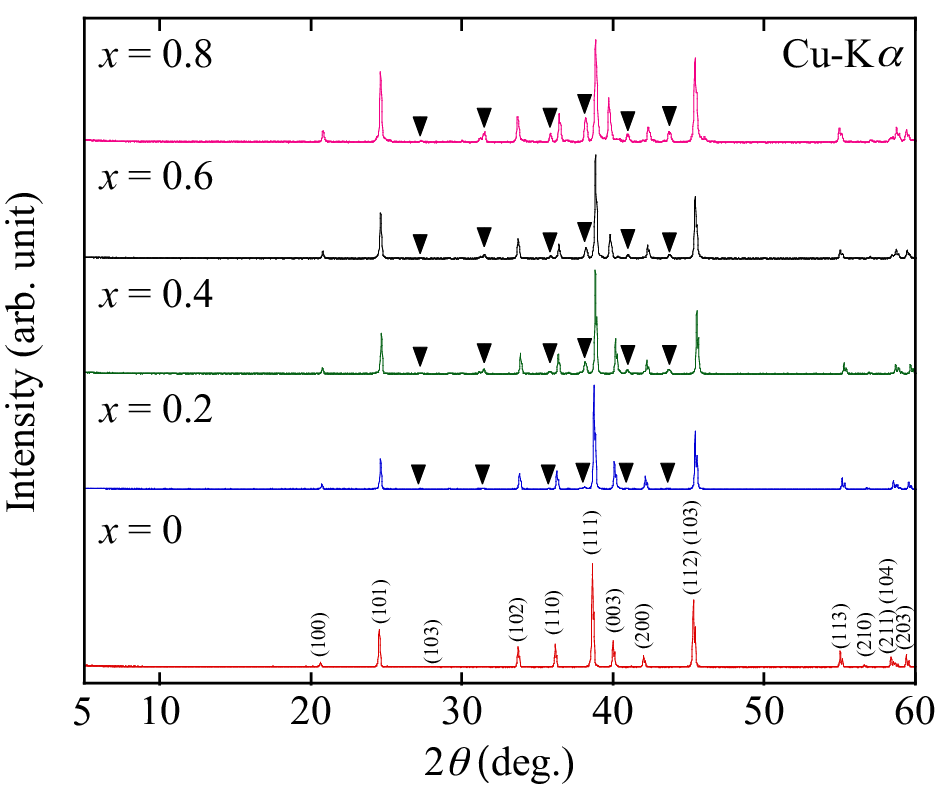}
\end{center}
\caption{
Powder X-ray diffraction pattern of TaAl$_{x}$Ge$_{2-x}$, synthesized by the solid-state reaction.
Miller indices on the peaks in $x = 0$ indicate
Bragg reflections originated from chiral TaGe$_{2}$.
Inverse triangles on the peaks denote an impurity phase of Ta$_{5}$Ge$_{3}$.
}
\label{f-PXRD}
\end{figure}

Figure~\ref{f-PXRD} shows the powder X-ray diffraction pattern of the synthesized TaAl$_{x}$Ge$_{2-x}$ with $x$ from 0 to 0.8.
When $x$ is smaller than 0.2, a single phase of the chiral C40 compound is obtained as seen in the data for $x = 0$, where all the diffraction peaks are assigned to the indices of the C40 structure.
In the specimen with $x$ of 0.2, some extra peaks appear as indicated by inverse triangles,
indicating a presence of impurity phases of Ta$_5$Ge$_3$ associated with tetragonal Cr$_5$B$_3$-type (space group $I4/mcm$) and hexagonal Mn$_5$Si$_3$-type (space group $P6_{3}/mcm$) structures.
The weight fraction of such impurity phases is a few percent for $x = 0.2$. It increases with elevating $x$ and reaches 16.5\% in the specimen with $x$ of 0.8.

\begin{figure}[tb]
\begin{center}
\includegraphics[width=7.5cm]{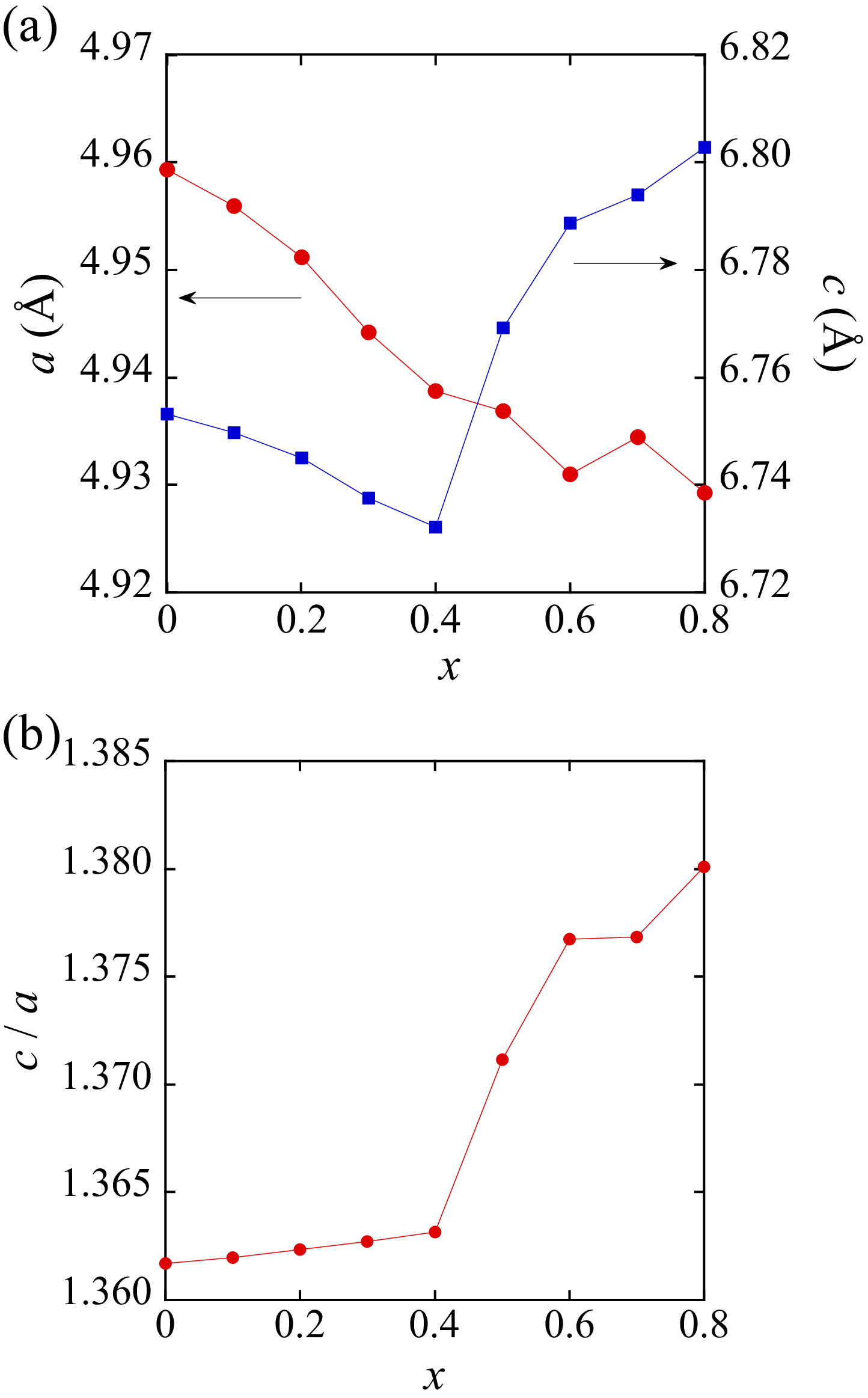}
\end{center}
\caption{
Lattice constants of TaAl$_{x}$Ge$_{2-x}$, deducted by powder X-ray diffraction data.
(a) the lattice constants $a$ and $c$, and (b) their ratio as a function of Al-substitution $x$.
}
\label{f-LatticeConst}
\end{figure}

Figure~\ref{f-LatticeConst} shows an evolution of the lattice constants $a$ and $c$ in the C40 structure of TaAl$_{x}$Ge$_{2-x}$.
They monotonously shrink with increasing $x$ from 0 to 0.4.
When exceeding 0.4, the $c$ drastically increases although the $a$ continues to decrease with further increasing $x$.
The ratio of the $c$ to the $a$, as shown in Fig.~\ref{f-LatticeConst}(b), takes almost a constant value in the region of $x$ from 0 to 0.4.
Then, it starts to expand significantly above 0.4 and increases by 1.5\% at $x$ = 0.8.

\begin{figure}[tb]
\begin{center}
\includegraphics[width=7.5cm]{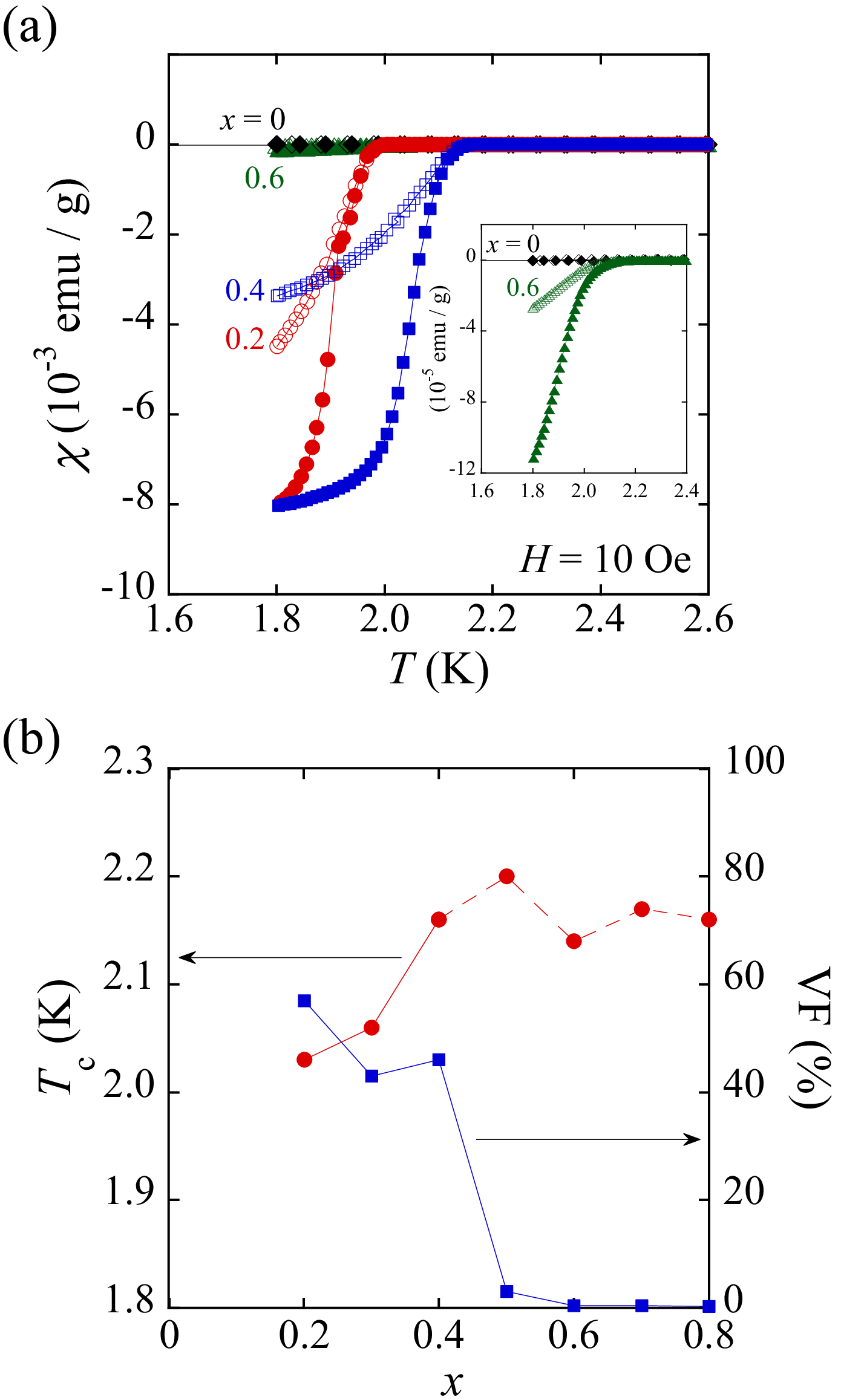}
\end{center}
\caption{(a) Temperature dependence of the magnetic susceptibility of TaAl$_{x}$Ge$_{2-x}$ under a magnetic field of 10 Oe.
Diamonds, circles, squares and triangles denote the susceptibility of the specimens with $x$ of 0, 0.2, 0.4 and 0.6, respectively.
Closed and open marks present the susceptibility collected by the zero-field cooling (ZFC) and field cooling (FC) processes, respectively.
The inset shows an enlarged view to see the small susceptibility data for the specimens with $x$ of 0 and 0.6.
(b) Al-content $x$ evolution of $T_{\rm c}$ and volume fraction (VF) of TaAl$_{x}$Ge$_{2-x}$.
}
\label{f-MT}
\end{figure}

Figure~\ref{f-MT}(a) shows a temperature dependence of the magnetic susceptibility of the synthesized TaAl$_{x}$Ge$_{2-x}$ with $x$ from 0 to 0.6.
It is clear that the Al substituted specimens with $x$ of 0.2 to 0.4 show superconductivity. They exhibit a sharp drop of the susceptibility both in the ZFC and FC processes at 2.03 and \SI{2.16}{\kelvin} for $x = 0.2$ and 0.4, respectively.
The superconducting VF in the FC process at 1.8 K is estimated to be 71 and 53\% for $x = 0.2$ and 0.4, respectively.
These values are large enough to identify the superconductivity in TaAl$_{x}$Ge$_{2-x}$.
However, our single-phase bulk specimens of TaGe$_{2}$ do not show any signal of superconductivity down to \SI{1.8}{\kelvin}.
This result is not consistent with that reported in the previous study \cite{Knoedler1979},
in which the bulk specimen of TaGe$_{2}$ showed superconductivity at \SI{1.9}{\kelvin} in the resistance measurement.

Note that the superconductivity appears at \SI{2.14}{\kelvin} in the ZFC process of susceptibility for $x = 0.6$, as shown in the inset of Fig.~\ref{f-MT}(a).
However, the VF is estimated to be 0.6\% for $x = 0.6$.
It also remains very small for larger $x$, as shown in Fig. 4(b), suggesting that the superconductivity is significantly suppressed when $x$ is larger than 0.4.
Considering the features observed above 0.4; almost the same values of $T_{c}$ as that for $x$ of 0.4, the small VF, and the sudden increase of the lattice constant $c$,
it is natural to regard that the Al substitution occurs inhomogeneously in the specimen in this regime.
Namely, most part of the specimen is successfully substituted with the expected Al content but does not show superconductivity,
while a small part of the specimen does show it because it contains an appropriate Al content (smaller than 0.4).

\begin{figure}[tb]
\begin{center}
\includegraphics[width=6.8cm]{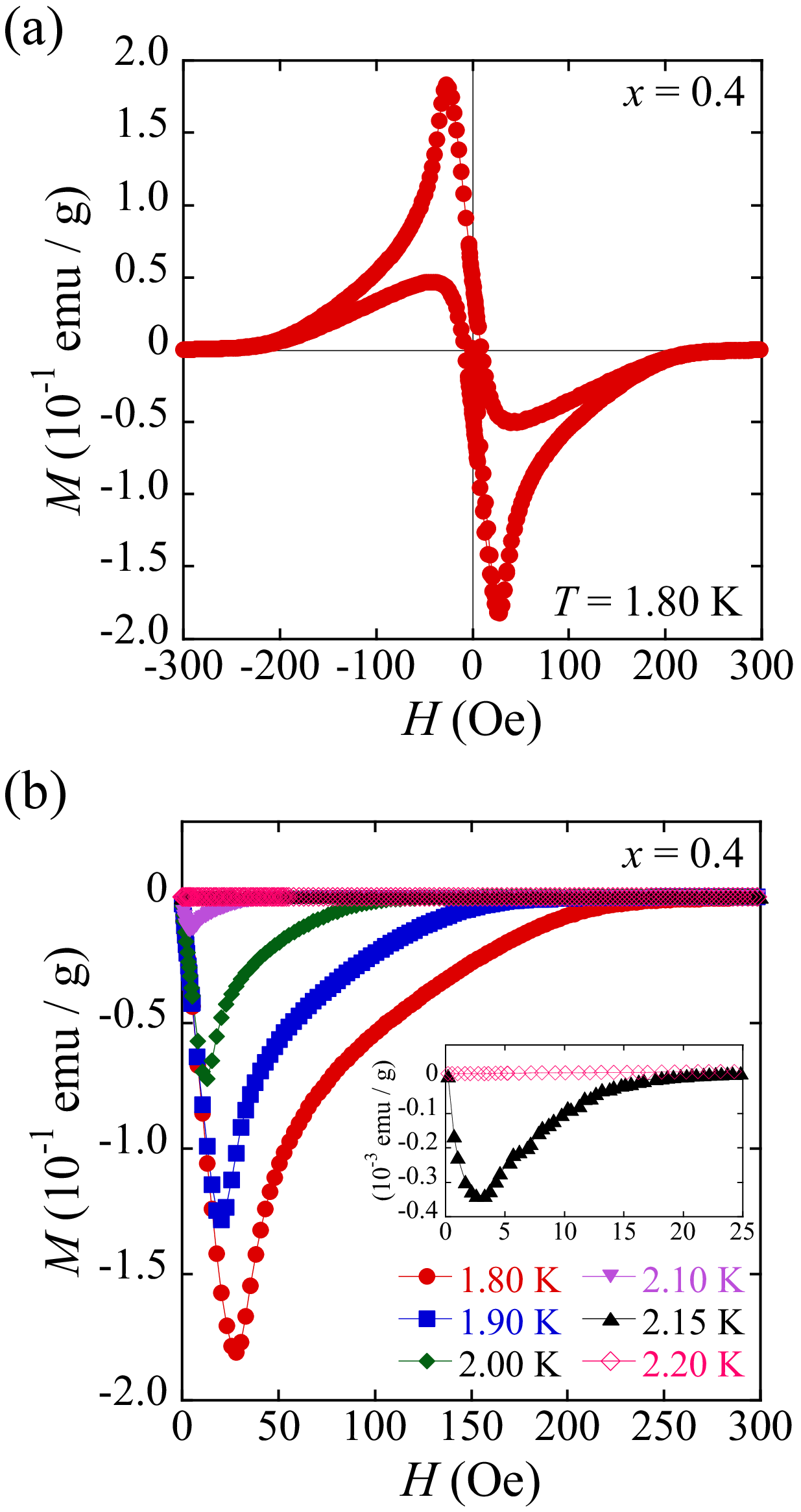}
\end{center}
\caption{
Magnetic field dependence of the magnetization of TaAl$_{0.4}$Ge$_{1.6}$.
(a) A hysteresis full loop of the $M$-$H$ curve collected at \SI{1.80}{\kelvin}.
(b) $M$-$H$ curves at various temperatures.
The inset shows the enlarged view of the low $H$ region at 2.15 and \SI{2.20}{\kelvin}.
}
\label{f-MH}
\end{figure}

Figure~\ref{f-MH} shows a magnetic field dependence of the magnetization of TaAl$_{0.4}$Ge$_{1.6}$.
A typical behavior of the type-II superconductivity appears in a hysteresis full loop of the $M$-$H$ curve in Fig.~\ref{f-MH}(a).
The $H_{\rm c1}(T)$ and the $H_{\rm c2}(T)$ are deducted from the $M$-$H$ curves at various temperatures in Fig.~\ref{f-MH}(b).
Here, the $H_{\rm c1}(T)$ is defined as the field deviated from the initial slope of linear diamagnetic behavior,
and the $H_{\rm c2}(T)$ is defined as the field where the hysteresis loop closes.
Similar behavior of the type-II superconductivity was observed in the case of TaAl$_{0.2}$Ge$_{1.8}$ with smaller values of $H_{\rm c1}$ and $H_{\rm c2}$.

\begin{figure}[tb]
\begin{minipage}[b]{1.0\linewidth}
  \centering
  \includegraphics[width=6.8cm]{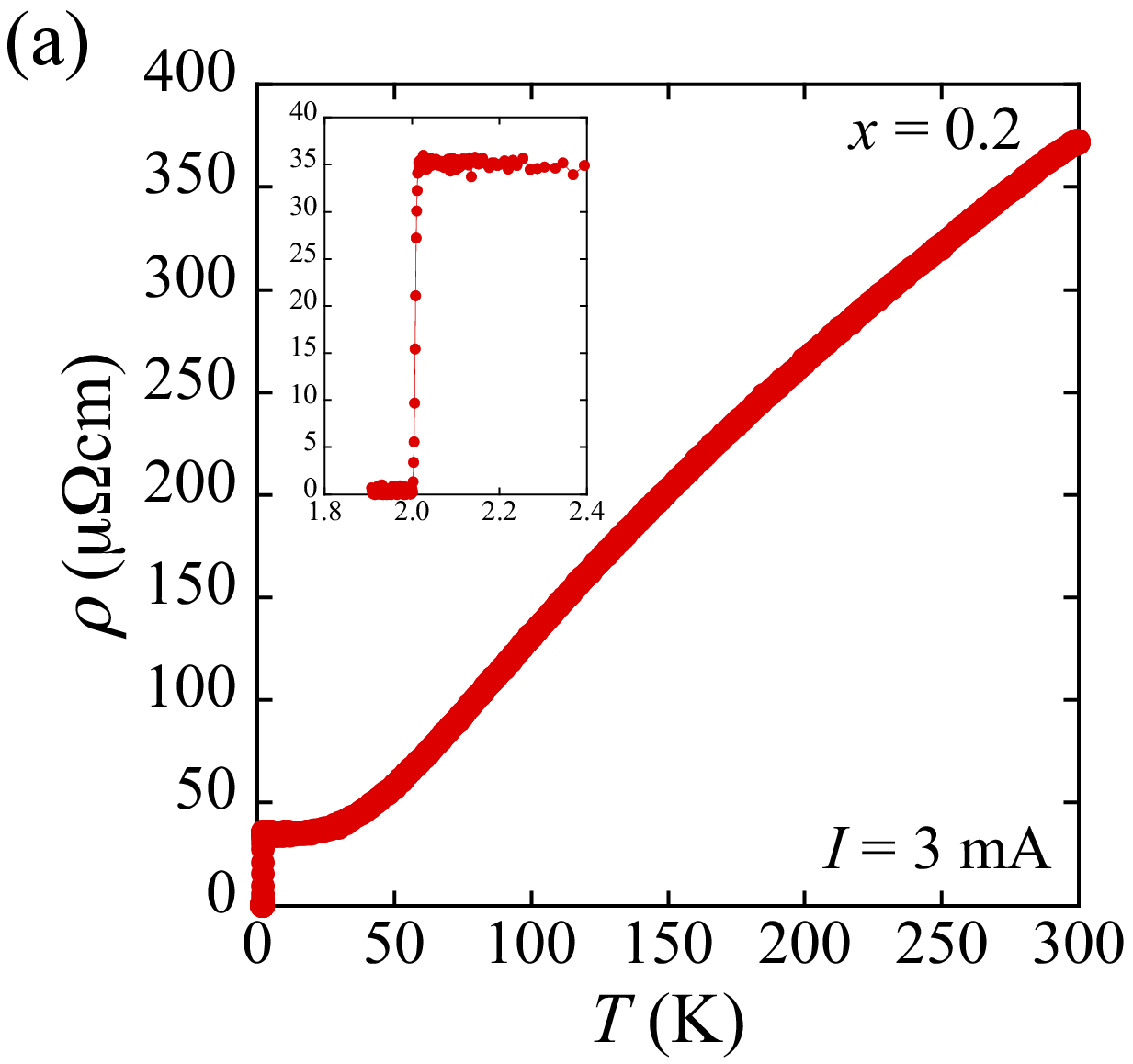}
\end{minipage}
\begin{minipage}[b]{1.0\linewidth}
  \centering
  \includegraphics[width=6.8cm]{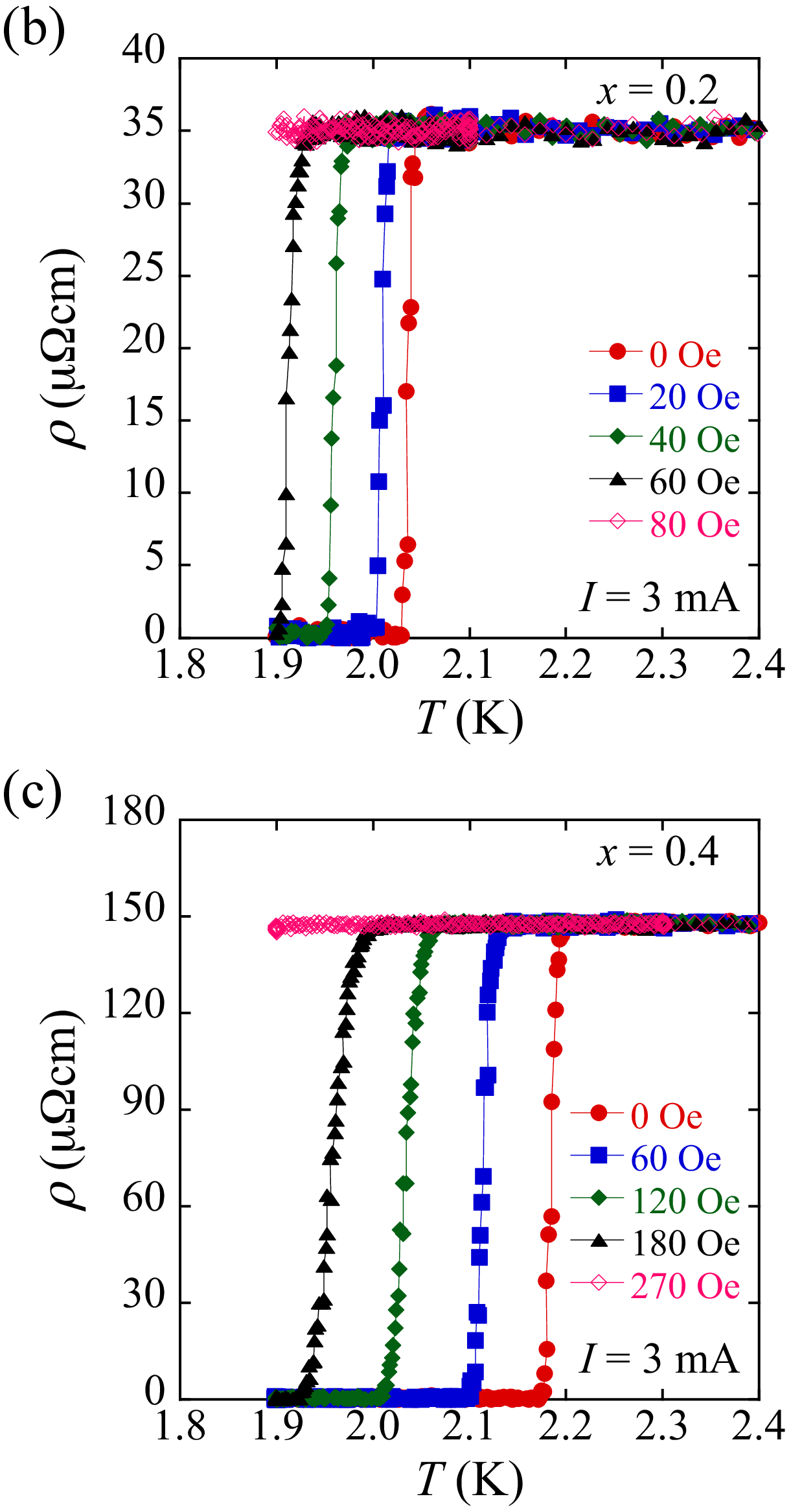}
\end{minipage}
\caption{
Temperature dependence of the electrical resistivity of TaAl$_{0.2}$Ge$_{1.8}$ and TaAl$_{0.4}$Ge$_{1.6}$.
The current value is fixed to be \SI{3}{\milli \ampere}.
(a) The data at zero magnetic field from \SI{300}{\kelvin} to the lowest temperature for $x = 0.2$, with an enlarged view of the low $T$ region.
(b) and (c) The data at various magnetic fields in low temperature region for $x = 0.2$ and 0.4, respectively.
}
\label{f-RT}
\end{figure}

Figure~\ref{f-RT} shows a temperature dependence of the electrical resistivity of TaAl$_{x}$Ge$_{2-x}$ with $x = 0.2$ and 0.4.
The resistivity sharply drops at the onset temperature $T_{\rm{onset}}$ of \SI{2.04}{\kelvin},
and then reaches to zero at $T_{\rm{zero}}$ of \SI{2.03}{\kelvin} at zero magnetic field in TaAl$_{0.2}$Ge$_{1.8}$, as shown in Fig.~\ref{f-RT}(a).
The residual resistivity ratio between \SI{300}{\kelvin} and $T_{\rm{onset}}$ is 11.
Similar behavior of the superconducting phase transition is observed in TaAl$_{0.4}$Ge$_{1.6}$ with $T_{\rm{onset}}$ of \SI{2.20}{\kelvin} and $T_{\rm{zero}}$ of \SI{2.17}{\kelvin}.

Figures~\ref{f-RT}(b) and \ref{f-RT}(c) show the superconducting phase transition in the presence of magnetic fields in TaAl$_{0.2}$Ge$_{1.8}$ and TaAl$_{0.4}$Ge$_{1.6}$, respectively.
In both specimens, $T_{\rm c}$ systematically decreases with increasing the strength of magnetic field.
This behavior provides a series of $H_{\rm c2}(T)$, defined by the middle-point temperature of the transition at each magnetic field.

\begin{figure}[tb]
\begin{center}
\includegraphics[width=6.8cm]{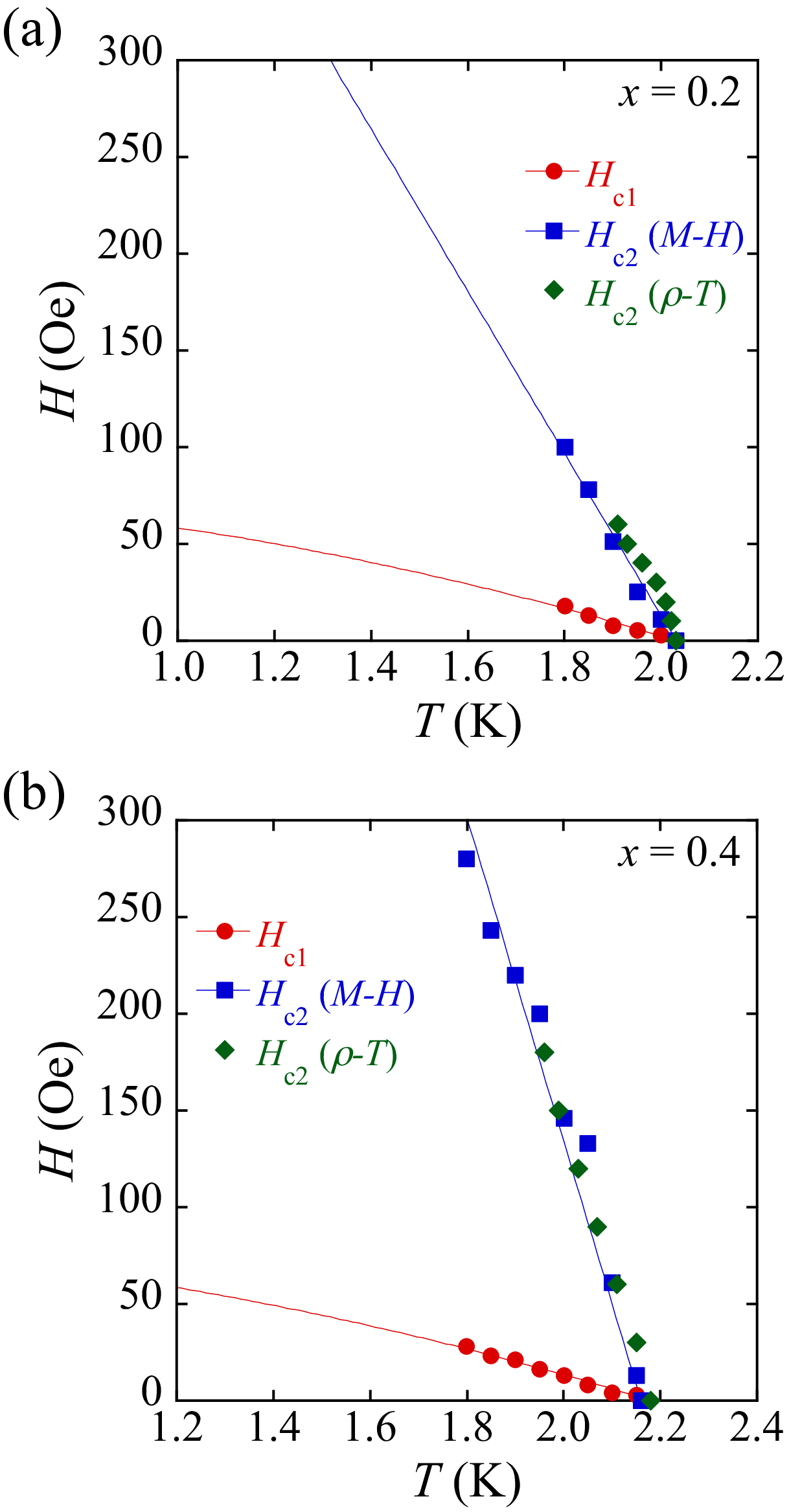}
\end{center}
\caption{
Magnetic field-temperature ($H$-$T$) phase diagram for superconductivity in (a) TaAl$_{0.2}$Ge$_{1.8}$ and (b) TaAl$_{0.4}$Ge$_{1.6}$.
The $H_{c1}$ was deducted by $M$-$T$ measurements.
The $H_{c2}$ was deducted by $M$-$H$ and $\rho$-$T$ measurements, represented by blue squares and green diamonds. 
}
\label{f-HT}
\end{figure}

The superconducting phase diagrams of TaAl$_{0.2}$Ge$_{1.8}$ and TaAl$_{0.4}$Ge$_{1.6}$ are summarized in Fig.~\ref{f-HT}.
The $H_{\rm c1}(T)$ line is fitted by the following formula derived from the Ginzburg-Landau theory,
$H_{\rm c1}(T)=H_{\rm c1}(0)[1-(T/T_{\rm c})^2]$.
The $H_{\rm c1}(0)$ is determined to be $74.2$ Oe and $89.3$ Oe for TaAl$_{0.2}$Ge$_{1.8}$ and TaAl$_{0.4}$Ge$_{1.6}$, respectively.
These values correspond to the penetration depths $\lambda_{\rm GL}$ of 298 and \SI{272}{\nano \metre} with a relation of $\mu_{0}H_{\rm{c1}}(0) \sim \Phi_{0}/\pi\lambda_{\rm GL}^{2}$,
where $\mu_{0}$ and $\Phi_{0}$ are the magnetic permeability of the vacuum and the quantum flux, respectively.
The $H_{\rm c2}$ line grows linearly with decreasing temperature down to \SI{1.8}{\kelvin}.
A linear extrapolation to \SI{0}{\kelvin} indicates the $H_{\rm c2}(0)$ is roughly estimated to be 850 Oe and 1810 Oe in TaAl$_{0.2}$Ge$_{1.8}$ and TaAl$_{0.4}$Ge$_{1.6}$, respectively.
These values of $H_{\rm c2}(0)$ are significantly smaller than the Pauli paramagnetic limit $H_{\rm{P}} $,
given as $18.4 \, T_{\rm c}$ (kOe) \cite{Chandrasekhar1962, Clogston1962}.
The corresponding coherent lengths $\xi_{\rm GL}$ are 88.0 and \SI{60.3}{\nano \metre}, giving $\kappa_{\rm GL}$
of 3.4 and 4.5 for $x$ = 0.2 and 0.4, respectively.
The values obtained for $\kappa_{\rm GL}$ are much larger than that in NbGe$_{2}$ \cite{Lv2020}
and support the type-II nature in the TaAl$_{x}$Ge$_{2-x}$ compounds.
To make further discussion of the superconducting state concerning the precise values of $H_{\rm c2}(0)$ and superconducting properties
uniquely observed in non-centrosymmetric and chiral crystals, additional measurements with lower temperature will be necessary.

\section{Discussions}

Let us discuss characteristics experimentally obtained in the C40-type TaGe$_{2}$ and its Al substituted compounds.

Our single-phase bulk specimens of TaGe$_{2}$ did not show superconductivity down to \SI{1.8}{\kelvin}. 
This behavior is not consistent with the previous report \cite{Knoedler1979}, where bulk specimens with $T_{\rm c}$ of \SI{1.9}{\kelvin} were synthesized by using an arc-melting method, whereas thin films with $T_{\rm c}$ of \SI{2.7}{\kelvin} were obtained via sputtering using the bulk specimens as a target.
The phase identification was performed only in the thin films using X-ray diffraction, resulting in an existence of some impurity phases. 
Thus, the bulk specimens were likely to contain such impurity phases, which might exhibit superconducting behavior.
Moreover, first-principle calculations of TaGe$_{2}$ and NbGe$_{2}$ indicated that
the $T_{\rm c}$ value in TaGe$_{2}$ should be lower than that in NbGe$_{2}$ through an estimation of Debye temperature and DOS at the Fermi level \cite{Kabir2023}.
To examine this possibility, further experiments at lower temperature will be required.

Superconductivity appears clearly in TaAl$_{x}$Ge$_{2-x}$ with $x$ of 0.2 to 0.4.  A monotonous decrease of the lattice constants $a$ and $c$ with increasing Al substitution $x$ up to 0.4 is in accord with the rise of $T_{\rm c}$ value from 2.0 to \SI{2.2}{\kelvin} in that regime.
First-principle calculations of TaGe$_{2}$ demonstrated that an orbital hybridization among Ta and Ge occurs close to the Fermi level \cite{Kabir2023}.
In this respect, the DOS can be easily affected by substituting Ge with Al and may contribute to an emergence of superconductivity in TaAl$_{x}$Ge$_{2-x}$.

$M$-$H$ curves show a type-II nature of superconductivity in TaAl$_{x}$Ge$_{2-x}$. 
This behavior is in strong contrast with a type-I nature observed in TaSi$_{2}$, another chiral superconductor of C40 compounds \cite{Gottlieb1992}.
In the phase diagram, the $H_{\rm c2}$(0) values, extracted from a linear extrapolation of the $H_{\rm c2}(T)$ line from the vicinity of $T_{\rm c}$,  become double with increasing $x$ from 0.2 to 0.4, while the $T_{c}$ value increases just by 10\%.
This behavior may suggest that the spin-triplet component is enhanced in the mixed Cooper pairing state with increasing the Al substitution, although the $H_{c2}$(0) value still remains much smaller than $H_{\rm P}$.

When $x$ reaches and exceeds 0.5,
a sudden disappearance or suppression of superconductivity occurs accompanying a rapid increase of the lattice constant $c$ and a continuous decrease of the lattice constant $a$. 
The former behavior may suggest an ordered Al substitution in the present compounds, leading to a possible formation of superstructure without occupying the same Wyckoff position as that for Ge.
Indeed, such a correlation of the lattice constants and superconductivity has been argued in $n$H-CaAlSi \cite{Sagayama2006,Kuroiwa2006}, which forms AlB$_{2}$-based multi-stack structures with $n$ of 1, 5 and 6.
Interestingly, superstructures of $n$H-CaAlSi were detected as diffraction spots with tiny intensity by means of synchrotron X-ray radiation.
In this respect, it is worth investigating the presence of the superlattice in TaAl$_{x}$Ge$_{2-x}$ by using synchrotron X-ray.

Another interesting view on the data is that the expansion of the lattice $c$ may trigger the suppression of superconductivity with $x$ of 0.5 or more.
This feature means that the superconductivity may recover with a uniaxial pressure applied along the $c$-axis in this regime.

\section{Concluding Remarks}

A series of Al substituted TaGe polycrystals were synthesized by means of the solid-state reaction.
A type-II superconductivity was observed in the TaAl$_{x}$Ge$_{2-x}$ samples of $x$ = 0.2 to 0.4 with $T_{\rm c}$ = 2.0 to \SI{2.2}{\kelvin}, respectively.

The present results suggest that the C40-type chiral disilicides and digermanides $TX_{2}$ are a promising family of materials for exploring chiral superconductors.
These compounds have many advantages in terms of chirality-controlled crystal growth \cite{Kousaka2023}
and chirality-induced physical properties \cite{Shiota2021,Shishido2021,Shishido2023}.
In particular, the chirality control was successfully demonstrated in crystal growth of NbSi$_2$ and TaSi$_2$ using a floating zone method with composition-gradient feed rods. The chirality-dependent spin polarization appears in such crystals with current injection. Thus, the spin-polarized state via the CISS may be in favor of enhancing the spin-triplet component in the mixed Copper pairing state in chiral superconductors, leading to a current-driven enhancement of superconducting properties.

\ack

We appreciate a fruitful discussion with K. Horigane concerning the phase identification in powder X-ray diffractograms.
This work was supported by JSPS KAKENHI Grant Numbers 21H01032, 22H01944, 23H01870 and 23H00091.

\end{document}